\documentstyle[11pt,fleqn]{article}
\catcode`\@=11
\@addtoreset{equation}{section}
\def\theequation{\arabic{section}.\arabic{equation}}
\def\appendix{\renewcommand{\thesection}{\Alph{section}}\setcounter{section}{0}
              \renewcommand{\theequation}
            {\mbox{\Alph{section}.\arabic{equation}}}\setcounter{equation}{0}}
\def\maketitle{\thispagestyle{empty}\setcounter{page}0\newpage
                \renewcommand{\thefootnote}{\arabic{footnote}}
                  \setcounter{footnote}0}
\renewcommand{\thanks}[1]{\renewcommand{\thefootnote}{\fnsymbol{footnote}}
               \footnote{#1}\renewcommand{\thefootnote}{\arabic{footnote}}}
\newcommand{\preprint}[1]{\hfill{\sl preprint - #1}\par\bigskip\par\rm}
\renewcommand{\title}[1]{\begin{center}\Large\bf #1\end{center}\rm\par\bigskip}
\renewcommand{\author}[1]{\begin{center}\Large #1\end{center}}
\newcommand{\address}[1]{\begin{center}\large #1\end{center}}
\def\dip{\smallskip Departimento de Fisica, Universidad del Valle,\\
                                 A.A. 25360, Cali, Colombia\\
                                 and\\
                                 Tomsk Pedagogical University,\\
                                 634041 Tomsk,Russia \\
                                 and\\
                                 Dept.of Physics, Hiroshima Univ.,\\
                                 Hiroshima, Japan }
\def\Idip{\address{\dip}}
\newcommand{\email}[1]{e-mail: \sl #1@quantum.univalle.edu.co\rm}
\newcommand{\femail}[1]{\thanks{\email{#1}}}
\newcommand{\pacs}[1]{\smallskip\noindent{\sl PACS numbers:
                       \hspace{0.3cm}#1}\par\bigskip\rm}
\def\babs{\hrule\par\begin{description}\item{Abstract: }\it}
\def\eabs{\par\end{description}\hrule\par\medskip\rm}
\renewcommand{\date}[1]{\par\bigskip\par\sl\hfill #1\par\medskip\par\rm}
\newcommand{\ack}[1]{\par\section*{Acknowledgments} #1}
\newcommand{\s}[1]{\section{#1}}

\def\M{{\cal M}}                       
\newtheorem{theorem}{Theorem}                  
\newtheorem{proposition}{Proposition}          
\def\R{{\hbox{{\rm I}\kern-.2em\hbox{\rm R}}}}   
\def\H{{\hbox{{\rm I}\kern-.2em\hbox{\rm H}}}}   
\def\N{{\hbox{{\rm I}\kern-.2em\hbox{\rm N}}}}   
\def\C{{\ \hbox{{\rm I}\kern-.6em\hbox{\bf C}}}} 
\def\Z{{\hbox{{\rm Z}\kern-.4em\hbox{\rm Z}}}}   
\def\Tr{\mathop{\rm Tr}\nolimits}                  
\def\Res{\mathop{\rm Res}\nolimits}                
\renewcommand{\Re}{\mathop{\rm Re}\nolimits}       

\def\citen#1{%
\edef\@tempa{\@ignspaftercomma,#1, \@end, }
\edef\@tempa{\expandafter\@ignendcommas\@tempa\@end}%
\if@filesw \immediate \write \@auxout {\string \citation {\@tempa}}\fi
\@tempcntb\m@ne \let\@h@ld\relax \let\@citea\@empty
\@for \@citeb:=\@tempa\do {\@cmpresscites}%
\@h@ld}
\def\@ignspaftercomma#1, {\ifx\@end#1\@empty\else
   #1,\expandafter\@ignspaftercomma\fi}
\def\@ignendcommas,#1,\@end{#1}
\def\@cmpresscites{%
 \expandafter\let \expandafter\@B@citeB \csname b@\@citeb \endcsname
 \ifx\@B@citeB\relax 
    \@h@ld\@citea\@tempcntb\m@ne{\bf ?}%
    \@warning {Citation `\@citeb ' on page \thepage \space undefined}%
 \else
    \@tempcnta\@tempcntb \advance\@tempcnta\@ne
    \setbox\z@\hbox\bgroup 
    \ifnum\z@<0\@B@citeB \relax
       \egroup \@tempcntb\@B@citeB \relax
       \else \egroup \@tempcntb\m@ne \fi
    \ifnum\@tempcnta=\@tempcntb 
       \ifx\@h@ld\relax 
          \edef \@h@ld{\@citea\@B@citeB}%
       \else 
          \edef\@h@ld{\hbox{--}\penalty\@highpenalty \@B@citeB}%
       \fi
    \else   
       \@h@ld \@citea \@B@citeB \let\@h@ld\relax
 \fi\fi%
 \let\@citea\@citepunct
}
\def\@citepunct{,\penalty\@highpenalty\hskip.13em plus.1em minus.1em}%
\def\@citex[#1]#2{\@cite{\citen{#2}}{#1}}%
\def\@cite#1#2{\leavevmode\unskip
  \ifnum\lastpenalty=\z@ \penalty\@highpenalty \fi 
  \ [{\multiply\@highpenalty 3 #1
      \if@tempswa,\penalty\@highpenalty\ #2\fi 
    }]\spacefactor\@m}
\begin{document}

\preprint{}
\title{
Statistical Entropy of Near-Extremal and Fundamental Black p-Branes}
\author{A. A. Bytsenko\thanks{email: abyts@spin.hop.stu.neva.ru}}
\address{State Technical University, St. Petersburg 195251, Russia}
\author{S. D.  Odintsov\femail{odintsov}}
\Idip
\date{}
\babs

The problem of asymptotic density of quantum states of fundamental extended
objects is revised in detail. We argue that in the near-extremal regime the
fundamental $p$-brane approach can yield a microscopic interpretation of the
black hole entropy. The asymptotic behavior of partition functions, associated
with the $p$-branes, and the near-extremal entropy of five-dimensional black
holes are explicitly calculated.

\eabs


\s{Introduction}

Semi-classical methods have revealed many intriguing thermodynamic
properties of black holes. In recent studies  these methods have been applied
 to the calculation of thermodynamic quantities for
black hole and extended objects in string theory or $M$-theory. As a result
the microscopic description of entropy in terms of the counting of string
states \cite{stro96-379-99,das96u-52,call96-475-645,horo96-77-2368,ghos96u-57,
beck96u-65,gubs96u-35,das96u-72,beck96-381-423,mald96-77-428,john96-378-78,
horo96-383-151,dijk96u-26,kleb96u-89} have been obtained. It has been shown that the
dilatonic scalar fields of the $p$-brane soliton have to be  finite on
the horizon in the near-extremal limit in order  for the tree-level approximation
to be reliable.

Note that isotropic extremal $p$-brane solutions of this type are very
limited. All of these cases were enumarated in Ref. \cite{lu96u-27}. Let us
start with  brief remarks on this list.

The elementary membrane \cite{duff91-253-113} and solitonic 5-brane in $D=11$
\cite{guve92-276-49}. For $D=11$ there is no dilaton field, and the microscopic
interpretation of the entropy of the membrane and $5$-brane was presented in
Ref. \cite{kleb96u-89}.

The self-dual 3-brane in $D=10$ \cite{horo91-360-197, duff91-273-409}. For
$D=10$ the dilaton field decouples from the self-dual $3$-brane; the
microscopic discussion of the entropy can be found in Ref. \cite{gubs96u-35}.

In the remaining cases the dilatonic scalar fields do not decouple for generic
values of charges, but they do remain finite at the horizon in the extremal
limit. When all charges are equal, the dilatonic scalar fields then decouple;
the resulting solutions are non-dilatonic $p$-branes \cite{kleb96u-89}. In this
case the dyonic string becomes the self-dual string in $D=6$
\cite{duff95-273-479}.

Black holes in $D=5$ and $D=4$ with three and four independent participating
field strengths respectively \cite{lu95u-53}. In these cases the black holes
become the Reissner-Nordstr\o{m} black holes \cite{haly96u-68}.
For $D=5$ black hole solution is stainless \cite{lu95-456-669}, but can be
oxidised to a boosted dyonic string in $D=6$ dimensions. It was shown that
the entropy per unit $p$-volume is preserved under the oxidisation and can be
associated with the microscopic counting of the corresponding $D$-string states
in the near-extremal regime \cite{stro96-379-99}.

In $D=4$ there are two regular black holes. One of them involves four field
strengths, and reduces to the Reissner-Nordstr\o{m} black hole when all the
charges are equal \cite{mald96-77-428,john96-378-78}. This is the case when
multiply-charged black holes (see Ref. \cite{cvet96-53-584}) may be regarded as
bound states at threshold of singly-charged black holes \cite{duff95-345-441,
sen95-440-421,sen95-10-2081,duff96-459-125,rahm95u-89}. The other involves
only two field strengths, each with electric and magnetic charges. This
solution was refered to as the dyonic black hole of the second type in
\cite{lu96-465-127}. The multi-center solutions of charged, dilatonic and
non-extremal black holes in $D=4$ can be found also in \cite{lu96u-26}.

In all these cases the number $N$ of non-vanishing charges compatible with
having some preserved supersymmetry has to be maximal. The non-extremal
generalization of these solutions were presented in Ref. \cite{duff96u-52}.
Recently the cosmological solutions in string and $M$-theory have been analysed
in \cite{lu96u-07,luka96u-38}.

There are many more $p$-brane solitons in the string or $M$-theory in addition
to the cases mentioned above. But at the classical level the ideal-gas relation
between entropy and temperature breaks down in all these other cases.
It happens due to the fact that for all these solitons the dilaton, as well
as the curvature and field strengths diverges on the horizon in the near-
extremal limit. The divergences may indicate a breakdown of validity of the
classical approximation. The inclusion of string and worldsheet loop
corrections can remove such singularities \cite{lu96u-27}.

It has been shown that in the non-dilatonic cases, the entropy and temperature
satisfy the massless ideal-gas relation $S\sim T^p$ \cite{kleb96u-89}, which
is in concordance with the microscopic $D$-brane picture. This analysis can
be extended to the regular dilatonic examples \cite{lu96u-27}.

The purpose of the present paper is to resolve these problems, comparing the
statistical properties of quantum states of near-extremal black and
fundamental $p$-branes.

The contents of the paper are the following. In Section 2 we summarize a
description of scalar $p$-brane classes in various dimensions. The statistical
entropy associated with some interacting fundamental $p$-brane exitation modes
and its comparison to the Bekenstein-Hawking and near-extremal black $p$-branes
entropy is given in Section 3. In Section 4 the explicit form of quantum
counting string states and the entropy of five-branes is computed. Finally
the Appendices contain explicit results for fundamental (super) $p$-branes,
namely:  mass operators (Appendix A), asymptotic expansions of
generating functions (Appendix B), and the one-loop free energy (Appendix C).

\s{Classes of Scalar $p$-Branes}

   In this section we start with the properties of classical $p$-branes.
Let us consider the relevant part of the tree-level approximation of
the string effective action. The bosonic Lagrangian has the form \cite
{lu95u-53,lu96-465-127,duff96u-52}
$$
e^{-1}{\cal L} = R-\frac{1}{2}\left(\partial \vec \phi \right)^2-\frac{1}{2n!}
\sum_{\alpha=1}^N e^{-\vec {a}_\alpha \cdot \vec \phi} \left(F^\alpha
\right)^2\mbox{,}
\eqno{(2.1)}
$$
where $\vec \phi =\left(\phi_1,...,\phi_N \right)$ is a set of $N$ scalar
fields, and $F^{\alpha}$ is a set of $N$ antisymmetric field strengths of rank
$n$, which give rise to a $p$-brane with world volume dimension $d=n-1$ if they
carry electric charges, or with $d=D-n-1$ if they carry magnetic charges. The
constant (dilatonic) vectors $\vec a_\alpha$ are vectors characteristic of
the supergravity theory associated with low energy limit of string or $M$-
theory.

The usual form of the metric is given by
$$
ds^2 = e^{2A(r)}dx^{\mu}dx^{\nu}\eta_{\mu\nu}+e^{2B(r)}dy^mdy^m\mbox{,}
\eqno{(2.2)}
$$
where $x^\mu (\mu=0,...,d-1)$ and $y^m$ are the coordinates of the $(d-1)$-
dimensional brane volume and $(D-d)$-dimensional transverse space
respectively. The functions $A(r)$, $B(r)$ and the dilatonic scalar
$\vec \phi$ depend on $r=\left(y^my^m\right)^{1/2}$ only. Note that the
metric anzatz therefore preserves an $SO(1,d-1)\bigotimes SO(D-d)$ subgroup
of the original $SO(1,D-1)$ Lorentz group.

For each $n$-index $(n\geq 2)$ field strengths $F^\alpha$ there are two
different equations that also preserve the same subgroup \cite{dabh90-340-33,
call91-359-611,call91-367-60}:
$$
F_{m\mu_1...\mu_{n-1}}^\alpha=\epsilon_{\mu_1...\mu_{n-1}}\left(e^{C_
\alpha(r)}\right)'\frac{y^m}{r}\mbox{,}
\eqno{(2.3)}
$$
$$
F_{m_1...m_n}^\alpha = \lambda_{\alpha}\epsilon_{m_1...m_np}
\frac{y^p}{r^{n+1}}\mbox{.}
\eqno{(2.4)}
$$
Here $C_\alpha(r)$ is a some differentiable function, $\epsilon_\mu$ is the
volume form on the unit sphere whose metric is $d\Omega^2$, and a prime
denotes a derivative with respect to $r$. The Eq. (2.3) gives rise to an
elementary $(d-1)$-dimensional brane with $d=n-1$, $n=2,3,4$ and electric
charges $\lambda_\alpha$. While for a solitonic $(d-1)$-dimensional brane
with $d=D-n-1$, $n=1,2,3,4$ and magnetic charges $\lambda_\alpha$ the second
Eq. (2.4) holds.

It should be noted that for $D=2n$ some field strengths $F^\alpha$ might be
the duals of the original ones of the same degree. Such a particularly
interesting class of solutions having both elementary and solitonic
contributions refers as dyonic solutions of the first type \cite{lu96-465-127}
which are possible only for $n=2$ $(D=4)$.

If the dot products $\M_{\alpha\beta}=\vec a_\alpha \cdot \vec a_\beta$
satisfy
$$
\M_{\alpha\beta}=4\delta_{\alpha\beta}-\frac{2d\tilde{d}}{2(D-2)}\mbox{,}
\eqno{(2.5)}
$$
where $\tilde{d}=D-d-2$, then the Lagrangian (2.1) can be embedded into the
string or $M$-theory \cite{lu95u-53}.

For the black $p$-branes with $N$ non-vanishing charges the metric has the form
\cite{duff96u-52}
$$
ds^2=e^{2A(r)}\left(e^{-2f}dt^2+dx^idx^i\right)+e^{2B(r)}\left(e^{-2f}dr^2+
r^2d\Omega^2\right)\mbox{,}
\eqno{(2.6)}
$$
where
$$
\exp(2A(r))=\prod_{\alpha=1}^N\left(1+\frac{k}{r^{\tilde{d}}}\sinh^2
\mu_\alpha\right)^{-\tilde{d}/(D-2)}\mbox{,}
\eqno{(2.7)}
$$
$$
\exp(2B(r))=\prod_{\alpha=1}^N\left(1+\frac{k}{r^{\tilde{d}}}\sinh^2
\mu_\alpha\right)^{d/(D-2)}\mbox{.}
\eqno{(2.8)}
$$
The $d$-dimensional world volume of the p-brane is parametrised by the
coordinates $(t,x^i)$, while the remaining coordinates are $r$ and the
coordinates on the $(D-d-1)$-dimensional unit sphere. It can be shown that
the function $f(r)$ has a completely universal form \cite{duff96u-52},
namely $\exp(2f(r))=1-kr^{-\tilde d}$. The dilatonic scalar fields
$\varphi_\alpha=\vec a_{\alpha}\cdot \vec \phi$ can be given by
$$
\exp(-\frac{1}{2}\epsilon\varphi_\alpha)=\left(1+\frac{k}{r^{\tilde{d}}}
\sinh^2\mu_\alpha\right)\prod_{\beta=1}^N\left(1+\frac{k}{r^{\tilde{d}}}
\sinh^2\mu_\beta\right)^{-\frac{d\tilde{d}}{2(D-2)}}\mbox{,}
\eqno{(2.9)}
$$
where $\epsilon=1$ $(-1)$ for the elementary (solitonic) solutions.
The metric of Eqs. (2.6)-(2.8) has a curvature singularity at $r=r_{-}=0$ and
an outer horizon at $r=r_{+}\equiv k^{1/\tilde{d}}$. Generally speaking
even at the origin in the extremal limit $r_{+}\mapsto r_{-}=0$ the
curvature remains singular. Futhermore the charges for each field strength
are given by $\lambda_\alpha=\frac{1}{2}\tilde{d}k\sinh 2\mu_\alpha$.
In the extremal limit we should take $k\mapsto 0$, $\mu_\infty \mapsto
\infty$ and keep the charges $\lambda_\alpha$ fixed.

When all charges are equal the Lagrangian (2.1) reduces to the functional which
describes a single scalar $p$-brane and a single field strength. In general
case for the non-singular matrix $\M_{\alpha\beta}$ the constant $a$, field
$\phi$ and strength $F$ are given by \cite{lu96-465-127}
$$
a^2=\left(\sum_{\alpha,\beta}\left(\M^{-1}\right)_{\alpha\beta}\right)^
{-1}\mbox{,}
\eqno{(2.10)}
$$
$$
\phi=a\sum_{\alpha,\beta}\left(\M^{-1}\right)_{\alpha\beta}\vec a_\alpha
\cdot\vec\phi\mbox{,}
\eqno{(2.11)}
$$
$$
\left(F^{\alpha}\right)^2=a^2\sum_\beta\left(\M^{-1}\right)_\beta^
\alpha F^2\mbox{.}
\eqno{(2.12)}
$$
The dilaton prefactor $a$ can be parametrised by $a^2=\triangle-2d\tilde{d}
(D-2)$. Supersymmetric solutions are associated to $\triangle=4/N$, with $N$
field strengths participating in the $p$-brane solution. In addition the
functions $A(r)$ and $B(r)$ can be written as follows \cite{duff96u-52}:
$$
\exp\left(2A(r)\right)=\left(1+\frac{k}{r^{\tilde{d}}}\sinh^2\mu
\right)^{-\frac{4\tilde{d}}{\triangle(D-2)}}\mbox{,}
\eqno{(2.13)}
$$
$$
\exp\left(2B(r)\right)=\exp\left(-2A(r)\frac{d}{\tilde{d}}\right)\mbox{.}
\eqno{(2.14)}
$$

In the case of $\triangle=4$, pure electric or pure magnetic black $p$-branes
have been considered in Ref. \cite{horo91-360-197} for $D=10$ dimensions and
in Ref. \cite{duff94-416-301} for $D\leq 11$ dimensions. The analysis was
generalized to other values of $\triangle$ in the case of near-extremal
$p$-branes in Refs. \cite{lu95-456-669,lu96-465-127}.

\s{Entropy of Near-Extremal Black and Fundamental $p$-Branes}

We first begin with the discussion of  the thermodynamic properties of classical $p$-
branes. The Hawking temperature and the entropy per unit $p$-volume of the
black $p$-brane are given by \cite{duff96u-52}
$$
T=\frac{\tilde{d}}{4\pi r_{+}}\prod_{\alpha=1}^N\left(\cosh
\mu_{\alpha}\right)^{-1}\mbox{,}
\eqno{(3.1)}
$$
$$
S=\frac{1}{4}r_{+}^{\tilde{d}+1}V(S^{\tilde{d}+1})\prod_{\alpha=1}^N
\cosh\mu_{\alpha}\mbox{,}
\eqno{(3.2)}
$$
where $V(S^{\tilde{d}+1})=2\pi^{\tilde{d}/2+1}\left[(\frac{1}{2}\tilde{d})!
\right]^{-1}$ is the volume of the unit $(\tilde{d}+1)$-dimensional sphere.
In the near-extremal limit, i.e. $k\ll\lambda_{\alpha}$ for all $\alpha$ the
relation between entropy and temperature takes the form
$$
S\approx\frac{\tilde{d}}{16\pi}V(S^{\tilde{d}+1})\left(16\pi^2
\tilde{d}^{-N-2}\right)^{\frac{\tilde{d}}{N\tilde{d}-2}}
\left(\prod_{\alpha=1}^N\lambda_{\alpha}\right)^
\frac{\tilde{d}}{N\tilde{d}-2}T^{\frac{2(\tilde{d}+1)-N
\tilde{d}}{N\tilde{d}-2}}\mbox{.}
\eqno{(3.3)}
$$
If
$$
N=\frac{2(D-2)}{d\tilde{d}}\mbox{,}
\eqno{(3.4)}
$$
then the entropy and temperature are related as follows
$$
S\approx\frac{\tilde{d}}{16\pi}V(S^{\tilde{d}+1})\left(16\pi^2
\tilde{d}^{-N-2}\right)^{\frac{d}{2}}
\left(\prod_{\alpha=1}^N\lambda_{\alpha}\right)^{\frac{d}{2}}
T^{d-1} \sim T^p\mbox{.}
\eqno{(3.5)}
$$
This dependence looks like the natural entropy of massless ideal gas predicted
by $D$-brane considerations. Indeed, open strings on a Dirichlet $p$-branes can
be analyzed as an ideal gas of massless objects in a $p$-dimensional space.
For the equal charge parameters $\lambda_{\alpha}$ the dilatonic
scalars decouple and the Eq. (3.5) reduces to the relation that is consistent
with the one found in Ref. \cite{kleb96u-89}. The above relation holds even
if the charges are not equal (the dilatonic scalar fields do not decouple but
remains finite at the horizon in the extremal limit) \cite{lu96u-27}. Note 
also that thermodinamic properties of black p-branes have been recently
discussed in ref.[61].

Note that in the near-extremal limit the curvature and the field strengths are
also finite at the horizon when the condition (3.4) is satisfied
\cite{lu96u-27}. For other values of $N$ the relation $S\sim T^{d-1}$ in the
near-extremal limit breaks down. It can be expected since dilatonic scalar
fields, the field strengths and the curvature diverge at the horizon in this
case. In fact the tree-level approximation is not sufficient for dilatonic
$p$-branes. But for loop effects, the dilaton fields, field strengths and
curvature may be finite at the horizon and, therefore, the relation between
entropy and temperature of the quantum $p$-branes will satisfy precisely the
natural ideal-gas scaling \cite{kleb96u-89,lu96u-27}.

In the near-extremal regime for all $p$-brane solitons satisfying the condition
(3.4) the temperature goes to zero. In this situation the relation between the
entropy and mass of the near-extremal $p$-branes can also be written as
$$
S\sim\left(\delta M^2\right)^{\frac{d-1}{d}}\mbox{.}
\eqno{(3.6)}
$$
On the other hand adapting to the thermodynamics of fundamental $p$-branes
$H_{\pm}(z)$ (see Eq. (B.1) in the Appendix B) can be regarded as a partition
function and $z\equiv\beta$ as the inverse temperature. Thus the related
statistical free energy ${\cal F}(\beta)$, entropy $S$ and internal energy $E$
may be written respectively as
$$
{\cal F}_{p}(\beta)=-\frac{1}{\beta}\mbox{log}\left[H_{-}(\beta)\right]\mbox{,}
\eqno{(3.7)}
$$
$$
{\cal F}_{sp}(\beta)=-\frac{1}{\beta}\mbox{log}\left[H_{+}(\beta)
H_{-}(\beta)\right]\mbox{,}
\eqno{(3.8)}
$$
$$
E=\frac{\partial}{\partial\beta}\left[\beta{\cal F}(\beta)\right]\mbox{,}
\eqno{(3.9)}
$$
$$
S=\beta^2\frac{\partial}{\partial\beta}{\cal F}(\beta)\mbox{.}
\eqno{(3.10)}
$$
Using  Eqs.(B.12) and (B.13) of the Theorem 1 one can obtain the
asymptotic density of (super) $p$-brane states in the form (see also
\cite{byts93-304-235,byts94-9-1569,byts96-266-1} for detail)
$$
\Omega(M)dM\simeq 2C_{\pm}(p)M^{\frac{2p-q}{2(1+p)}}
\exp\left[b_{\pm}(p)M^{\frac{2p}{1+p}}\right]dM\mbox{,}
\eqno{(3.11)}
$$
$$
b_{-}(p)\equiv\left(1+\frac{1}{p}\right)\left[q A\Gamma(1+p)
\zeta_{R}(1+p)\right]^{\frac{1}{1+p}}\mbox{,}
\eqno{(3.12)}
$$
$$
b_{+}(p)\equiv\left(1+\frac{1}{p}\right)\left[q A\Gamma(1+p)
(\zeta_{+}(1+p)+\zeta_{-}(1+p))\right]^{\frac{1}{1+p}}\mbox{,}
\eqno{(3.13)}
$$
where according to Eq. (B.9) we have $A=V(S^{p-1})$. The asymptotic density
is consistent with the entropy of near-extremal $p$-branes, indeed
$S\sim M^{\frac{2p}{p+1}}$, i.e. the relation (3.6) holds.
Note that asymptotic states density of compactified super p-branes which are
protected against usual topological instabilities has the same form as above
(and what is more the asymptotic behaviour of discrete states density has a
universal character for all p-branes \cite{byts96-266-1}).

In the limit $\beta\mapsto 0$ $(T\mapsto\infty)$ the entropy of fundamental
objects may be identified with $\mbox{log}(r_{\pm}(N))$, while the internal
energy is related to $N$. Hence from Eqs. (B.4),(B.5) and (3.7)-(3.10) one has
$$
{\cal F}_{p}(T)\simeq -qA\Gamma(p)\zeta_{R}(1+p)T^{p+1}\mbox{,}
\eqno{(3.14)}
$$
$$
{\cal F}_{sp}(T)\simeq -qA\Gamma(p)\left[\zeta_{+}(1+p)+\zeta_{-}(1+p)
\right]T^{p+1}\mbox{,}
\eqno{(3.15)}
$$
$$
E_{p}\simeq pq A\Gamma(p)\zeta_{R}(1+p)T^{p+1}\mbox{,}
\eqno{(3.16)}
$$
$$
E_{sp}\simeq pq A\Gamma(p)\left[\zeta_{+}(1+p)+\zeta_{-}(1+p)\right]
T^{p+1}\mbox{,}
\eqno{(3.17)}
$$
$$
S_{p}\simeq (1+p)q A\Gamma(p)\zeta_{R}(1+p)T^p\mbox{,}
\eqno{(3.18)}
$$
$$
S_{sp}\simeq (1+p)q A\Gamma(p)\left[\zeta_{+}(1+p)+\zeta_{-}(1+p)\right]
T^p\mbox{.}
\eqno{(3.19)}
$$
Eliminating the quantity $T$ between the Eqs. (3.16), (3.18) and (3.17),
(3.19) one gets
$$
S_{p}\simeq\frac{1+p}{p}\left[q pA\Gamma(p)\zeta_{R}(1+p)\right]^{\frac{1}{1+p}}
E_{p}^{\frac{p}{1+p}}\mbox{,}
\eqno{(3.20)}
$$
$$
S_{sp}\simeq\frac{1+p}{p}\left[q pA\Gamma\left(\zeta_{+}(1+p)+\zeta_{-}(1+p)
\right)\right]^{\frac{1}{1+p}}E_{sp}^{\frac{p}{1+p}}\mbox{.}
\eqno{(3.21)}
$$
Thus the entropy behavior can be understood in terms of the degeneracy of some
interacting fundamental $p$-brane exitation modes. Generally speaking the
fundamental $p$-brane approach can yield a microscopic interpretation of the
entropy.

It is well known that for $p>1$ the free energy power series in Eqs. (C.20) and
(C.21) (see Appendix C) diverge strongly for any $\beta$ (supermembranes free
energy at finite cut-off has been calculated in Refs. \cite{byod90}). There is
a conjecture that the critical temperature cannot be nonzero for $p>1$. This
argument has been suggested within the context of respective analysis of
$p$-brane thermodynamics \cite{alva91-43-3990,alva92-7-2889}. On the same time
the temperature goes to zero in the near-extremal limit for all $p$-brane
solitons satisfying the condition (3.4) and such solitons have vanishing
entropy in this limit as well.

More interesting possibility allows a finite temperature to be introduced into
the quantized fundamental (super) $p$-brane theory \cite{acto93-315-74}. The
proof of this statement is presented in the Appendix C. Our calculation does not
go through for even $p$, what suggests that there is some fundamental distinction
between even and odd $p$ in quantum theory of $p$-branes. Indeed the divergent
serie in Eqs. (C.20) and (C.21) for the odd $p$-branes free energy, when
reexpressed as ones on the left hand side of Eq. (C.28), remain well-defined for
finite temperature and have a smooth $\beta\mapsto\infty (T\mapsto 0)$ limit.
However it does prevent us from resumming the free energy, Eqs. (C.20) and
(C.21), for even $p$. Among other things, one can see a similar character of
non-dilatonic solutions of black branes as well. The even $p$ solutions are
singular while the odd $p$ solutions are not \cite{gibb95-12-297}. This may be
related to the fact that for even $p$ there cannot be a detailed agreement
between the Bekenstein-Hawking and statistical entropy \cite{kleb96u-89}.

\s{Near-Extremal Five-Branes and Black Hole Entropy}

First let us suppose that the second quantized theory of five-brane can be
considered as a theory of non-interacting strings. Then the Hilbert space of
all multiple string states that satisfy the BPS conditions (zero branes) with
a total energy momentum $P$ has the form \cite{dijk96u-26}
$$
{\cal H}_{P}=\bigoplus_{\scriptstyle \sum lN_{l}=N_{P}}\bigotimes_{l}Sym^
{N_{l}}{\cal H}_{l}\mbox{,}
\eqno{(4.1)}
$$
where symbol $Sym^N$ indicates the $N$-th symmetric tensor product. One can
expect that the exact dimension of ${\cal H}_{P}$ is determined by the character
expansion formula
$$
\sum_{N_{P}}dim{\cal H}_{P}q^{N_{P}}\simeq\prod_{l}\left(\frac{1+q^l}{1-q^l}
\right)^{\frac{1}{2}dim{\cal H}_l}\mbox{,}
\eqno{(4.2)}
$$
where the dimension ${\cal H}_l$ of the Hilbert space of single string BPS
states with momentum $k=l\hat{P}$ is given by
$dim{\cal H}_l=d(\frac{1}{2}l^2\hat{P})$, and $|\hat{P}|^2=|\hat{P}_{L}|^2-
|\hat{P}_{R}|^2$ (see Ref. \cite{dijk96u-26} for detail). The asymptotics
of the generating function (4.2) and the dimension ${\cal H}_{P}$ can be found
 with the help of Theorem 1 (Appendix B) that is generalization of the
Meinardus result for vector-valued functions.

The Eq. (4.2) is similar to the denominator formula of a (generalized) Kac-Moody
algebra \cite{borc95-120-161,harv96-463-315}. A denominator formula can be
written as follows
$$
\sum_{\sigma\in W}\left(sgn(\sigma)\right)e^{\sigma(\rho)}=
e^{\rho}\prod_{r>0}\left(1-e^r\right)^{mult(r)}\mbox{,}
\eqno{(4.3)}
$$
where $\rho$ is the Weyl vector, the sum on the left hand side is over all
elements of the Weyl group $W$, the product on the right side runs over all
positive roots (one has the usual notation of root spaces, positive roots,
simple roots and Weyl group, associated with Kac-Moody algebra) and each
term is weighted by the root multiplicity $mult(r)$. For the $su(2)$ level,
for example, an affine Lie algebra (4.3) is just the Jacobi triple product
identity. For generalized Kac-Moody algebras there is a denominator formula
$$
\sum_{\sigma\in W}\left(sgn(\sigma)\right)\sigma\left(e^{\rho}\sum_{r}
\epsilon(r)e^r\right)=e^{\rho}\prod_{r>0}\left(1-e^r\right)^{mult(r)}\mbox{,}
\eqno{(4.4)}$$
where the correction factor on the left hand side involves $\epsilon(r)$
which is $(-1)^n$ if $r$ is the sum of $n$ distinct pairwise orthogonal
imaginary roots and zero otherwise.

The Eq. (4.2) reduces to the standard superstring partition function for
$\hat{P}^2=0$ \cite{dijk96u-26}. The equivalent description of the second
quantized string states on the five-brane can be obtained by considering the
sigma model on the target space $\sum_{N}Sym^NT^4$. There is the correspondence
between the formula (4.1) and the term at order $q^{\frac{1}{2}N_p{\hat{P}}^2}$
in the expansion of the elliptic genus of the orbifold $Sym^{N_{P}}T^4$.
Using this correspondence one finds that the asymptotic growth is equal that of
states at level $\frac{1}{2}N_P{\hat{P}}^2$ in a unitary conformal field
theory with central charge proportional to $N_{P}$.

In conclusion let us consider the $D$-brane method that may be used for
calculation the ground state degeneracy of systems with quantum numbers of
certain BPS extreme black holes. A typical 5-dimensional example has been
analyzed in Refs. \cite{call96-475-645,mald96-475-679,haly96u-12}. Working in
the type $IIB$ string theory on $M^5\bigotimes T^5$ one can construct a $D$-
brane configuration such that the corresponding supergravity solutions describe
5-dimensional black holes. In this example five branes and one branes are
wrapped on $T^5$ and the system is given Kaluza-Klein momentum $N$ in one
of the directions. Therefore the three independent charges $(Q_1,Q_5,N)$ arise
in the theory, where $Q_1$, $Q_5$ are  electric and a magnetic charges
respectively (see Eqs. (2.3) and (2.4)). The naive $D$-brane picture gives
the entropy in terms of partition function $H_{\pm}(z)$ for a gas of
$Q_1Q_5$ species of massless quanta. For $p=1$ the integers $r_{\pm}$ in
Eqs. (B.12) and (B.13) represent the degeneracy of the state with momentum
$N$. Thus for $N\mapsto\infty$, using the Eqs. (B.12) and (B.13) of Theorem 1
one has
$$
\mbox{log}r_{\pm}(N)=\sqrt{q\zeta_{\pm}(2)N}-\frac{q+3}{4}\mbox{log}N
+ \mbox{log}C_{\pm}(1) + O(N^{-\kappa_{\pm}})\mbox{,}
\eqno{(4.5)}
$$
where
$$
C_{+}(1)=2^{-\frac{1}{2}}\left(\frac{q}{16}\right)^{\frac{q+1}{2}}\mbox{,}
\hspace{0.3cm} C_{-}(1)=C_{+}(1)\left(\frac{4}{3}\right)^{\frac{q+1}{2}}\mbox{.}
\eqno{(4.6)}
$$
For fixed $q=4Q_1Q_5$ the entropy is given by
$$
S_{sp}=\mbox{log}\left[r_{+}(N)r_{-}(N)\right]\simeq 2\pi\sqrt{Q_1Q_5N}
\left(\sqrt{\frac{2}{3}}+\sqrt{\frac{1}{2}}\right)-\left(\frac{3}{2}+
3Q_1Q_5\right)\mbox{log}N \mbox{.}
\eqno{(4.7)}
$$
This expression agrees with the classical black hole entropy.

Recently it has been pointed out that the classical result (4.7) is incorrect when the
black hole becomes massive enough for its Schwarzschild radius to exceed any
microscopic scale such as the compactification radii \cite{mald96-475-679,
haly96u-12}. Indeed, if the charges $(Q_1,Q_5,N)$ tend to infinity in fixed
proportion $Q_1Q_5=Q(N)$, then the correct formula does not agree with the
black hole entropy (4.7). If, for example, $Q(N)=N$, then using Eqs. (B.12) and
(B.13) for $N\mapsto\infty$ one finds $\mbox{log}\left[r_{+}(N)r_{-}(N)\right]
\sim N\mbox{log}N$. The naive D-brane prescription, therefore, fails to agree
with U-duality which requires symmetry among charges $(Q_1,Q_5,N)$
\cite{haly96u-12}.

\s{Conclusions}

In this paper we returned to the problem of asymptotic density of quantum
states for fundamental $p$-branes initiated in Refs. \cite{byts93-304-235,
acto93-315-74,byts94-9-1569}. We have shown that in the near-extremal regime
the fundamental $p$-brane approach can yield a microscopic interpretation of
the black hole entropy.

Indeed we realized a comparison between the asymptotic state density and the
entropy (3.18), (3.19) of fundamental $p$-branes and classical black holes
(3.3), (3.5). To this aim the explicit form of the total level brane density
$$
\Omega(M)\simeq {\cal C}_{\pm}(p,M)
\exp\left[b_{\pm}(p)M^{\frac{2p}{1+p}}\right]\mbox{,}
\eqno{(5.1)}
$$
where ${\cal C}_{\pm}(p,M)=2C_{\pm}(p)M^{\frac{2p-q}{2(1+p)}}$, has been
evaluated. Until our results the comparison between the statistical mechanical density
of states of black holes and branes is based on some part of the density, the
rest of it not being explicitly known. However, with the help of the
Meinardus theorem (Eqs. (B.12) and (B.13) of the Appendix B) we have computed
the complete $p$-brane state density (3.11), including the prefactors
${\cal C}_{\pm}(p,M)$ and the factors $b_{\pm}(p)$, depending on the dimension
of the embedding space. A prefactor for the degeneracy of black hole states
at mass level represents general quantum field corrections to the state
density and it has not been known before our calculation. Nevertheless an attempt was made to
compare the asymptotic state density of branes and related density of neutral 
black holes in Ref. \cite{byts94-9-1569}. In this paper we have shown that the
asymptotic behaviour of classical entropy of near-extremal black branes 
coincides  with the asymptotic degeneracy of some weakly 
interacting fundamental $p$-brane excitation modes.

Futher, using fundamental $p$-brane technology we have described and computed
 also the (near-extremal) entropy in the $D$-brane picture. The entropy is then
just the sum of (left- and right-moving) contributions. We find that the
remarkable Meinardus formulae can be used for the entropy calculation in a
$D$-brane inspired picture. In particular, the explicit computation of the 
black hole (and massive black hole) entropy in terms of independent five-brane
charges is given. At low energies and densities the statistical entropy (4.7) 
is in perfect agreement with results obtained in Ref. \cite{horo96-383-151}.

Thus we have evaluated the entropy by using classical black solutions and
$D$-brane picture. A detailed correspondence of these approaches can be
established with the help of fundamental $p$-brane partition function technique.
The picture we advocated above relies rather on odd $p$.
We hope that the methods used in the paper may shed light on the structure of
fundamental extended objects at finite temperature and origin of $p$-branes
entropy.

It has been demonstrated recently that BPS part of string spectrum for IIB
string compactified on a circle do match with BPS part
of supermembrane spectrum (see Refs.\cite{schwarz95,russo96}). In these
papers the same discrete spectrum for supermembrane has been used
as in the present work. This fact indicates that there are deep connections
between strings and membranes (at least they should be considered as
different corners of M-theory). Then different string results may be obtained
via membrane-string correspondence. Therefore even being no fundamental theory
the study of (super) p-branes may provide new deep insights in the
understanding of string theory and consistent formulation of M-theory.

\ack{We thank Profs. S. Zerbini and A.A. Actor for useful discussions.
This work was supported in part by Russian Universities grant No. 95-0-6.4-1
and in part by COLCIENCIES. The research of A.A.B. was supported in part by
Russian Foundation for Fundamental Research grant No. 95-02-03568-a.}

\s{Appendix A. (Super) p-Brane Mass Operator}

It is known that for the noncompactified extended objects the question of
reliability of the quasiclassical approximation is not absolutely clear
\cite{berg87-185-330,duff88-297-515}. Nevertheless even for compactified
$p$-branes the problem of the stability of classical solution is rather
complicated and the loop diagrams have to be calculated to solve it.

Quasiclassical quantization of fundamental (super) $p$-branes which propagate
in $D$-dimensional Minkowski space-time leads to the "number operators"
$N_{\vec n}^{(b,f)}$, with $\vec n = (n_1,...,n_p)\in \Z^{p}$, where $\Z$
is the ring of integer numbers. The operators $N_{\vec n}^{(b,f)}$ and the
(anti) commutation relations for the oscillators (operators in a Fock space)
can be found, for example, in Refs. \cite{berg87-185-330,duff88-297-515,
byts96-266-1}. The mass operators for the bosonic and supersymmetric $p$-branes
can be written respectively as follows
$$
M_{p}^2 = \sum_{i=1}^{D-p-1}\sum_{\vec n \in \Z^p/{\{0\}}}
\omega_{\vec n}N_{\vec n i}^{(b)}\mbox{,}
\eqno{(A.1)}
$$
$$
M_{sp}^2=\sum_{i=1}^{D-p-1}\sum_{\vec n \in \Z^p/{\{0\}}}
\omega_{\vec n}\left(N_{\vec n i}^{(b)}+N_{\vec n i}^{(f)}\right)\mbox{,}
\eqno{(A.2)}
$$
where the frequencies are given by
$$
\omega_{\vec n}^2 = \sum_{i=1}^p n_i^2\mbox{.}
\eqno{(A.3)}
$$

\s{Appendix B. Asymptotics for Generating Functions}

Let us consider multi-component versions of the classical generating
functions for partition functions, namely
$$
H_{\pm}(z)=\prod_{\vec n\in \Z^p/\{0\}}
\left[1\pm
\exp\left(-z\omega_{\vec n}\right)\right]^{\pm q}\mbox{,}
\eqno{(B.1)}
$$
where $z=y+2\pi ix$, $\Re z>0$, $q>0$ and $\omega_{\vec n}$ is given by
Eq. (A.3). The total number of quantum states can be described by the quantities
$r_{\pm}(N)$ defined by
$$
K_{\pm}(t)=\sum_{N=0}^{\infty}r_{\pm}(N)t^N \equiv H_{\pm}(-\log t)\mbox{,}
\eqno{(B.2)}
$$
where $t<1$, and $N$ is a total quantum number. The Laurent inversion formula
associated with the above definition takes the form
$$
r_{\pm}(N)=\frac{1}{2\pi i}\oint dt\frac{K_{\pm}(t)}{t^{N+1}}\mbox{,}
\eqno{(B.3)}
$$
where the contour integral is taken on a small circle about the origin.

We shall use the results of Meinardus \cite{mein54-59-338,mein54-61-289,
andr76b} that can be easily generalised to the vector-valued functions of the
(B.1) type (for more detail see Ref. \cite{byts96-266-1}).

\begin{proposition}

In the half-plane $\Re z>0$ there exists an asymptotic
expansion for $H_{\pm}(z)$ uniformly in $x$ as $y\mapsto 0$,
provided $|argz|\leq\frac{\pi}{4}$ and $|x|\leq\frac{1}{2}$ and given
by
$$
H_{+}(z)=\exp\left\{q[A\Gamma(p)\zeta_{-}(1+p)z^{-p}-Z_p(0)\mbox{log}2+
O\left(y^{c_{+}}\right)]\right\}\mbox{,}
\eqno{(B.4)}
$$
$$
H_{-}(z)=\exp\left\{q[A\Gamma(p)\zeta_{+}(1+p)z^{-p}-Z_p(0)\mbox{log}z+
Z_p^{'}(0)+O\left(y^{c_{-}}\right)]\right\}\mbox{,}
\eqno{(B.5)}
$$
where $0<c_{+},c_{-}<1$ and $Z_p(s)\equiv Z_p\left|_{\vec h}^{\vec g}
\right|(s)$ is the $p$-dimensional Epstein zeta function
$$
Z_p\left|_{\vec h}^{\vec g}\right|(s)\equiv \sum_{\vec n\in \Z^p}{'}\left(
\sum_{i=1}^p(n_i+g_i)^2\right)^{-s/2}\exp[2\pi i(\vec n,\vec h)]\mbox{,}
\eqno{(B.6)}
$$
which has a pole with residue $A$.
\end{proposition}

In above equations $\zeta_{-}(s)\equiv \zeta_R(s)$ is the Riemann zeta function,
$\zeta_{+}(s)=(1-2^{1-s})\zeta_{-}(s)$,\, $(\vec n,\vec h)=\sum_{i=1}^p
n_ih_i$,\, $g_i$ and $h_i$ are real numbers and the prime on $\sum {'}$
means to omit the term $\vec n =-\vec g$. For $\Re z<p$,\, $Z_p(s)$ is
understood to be the analytic continuation of the right hand side of the
Eq. (B.6). Futhermore $Z_p(s)$ is a fundamental zeta function which means it
has a functional equation
$$
Z_p\left|_{\vec h}^{\vec g}\right|(s)=\pi^{s-p/2}\frac{\Gamma(\frac{p-s}{2})}
{\Gamma(\frac{s}{2})}\exp[-2\pi(\vec g,\vec h)]Z_p\left|_{-\vec g}^{\vec h}
\right|(p-s)\mbox{.}
\eqno{(B.7)}
$$
Note that $Z_p(s)$ is an entire function in the complex $s$-plane except for the
case when all the $h_i$ are integers. In this case $Z_p(s)$ has a simple pole
at $s=p$,
$$
Z_p(p+\epsilon)=\frac{A}{\epsilon}B_p+O(\epsilon)\mbox{,}
\eqno{(B.8)}
$$
$$
A=\frac{2\pi^{p/2}}{\Gamma(\frac{p}{2})}\mbox{.}
\eqno{(B.9)}
$$
The constants $B_p$ can be evaluated by means of a recursion formula. Finally
from the functional equation (B.7) it follows that
$$
Z_p'(0)=\frac{B_p}{A}+\frac{1}{2}\psi(\frac{p}{2})-\mbox{log}\pi-
\frac{1}{2}\gamma\mbox{,}
\eqno{(B.10)}
$$
$$
Z_p(0)=-1\mbox{,}
\eqno{(B.11)}
$$
where $\psi(s)=\Gamma'(s)/\Gamma(s)$ and $\gamma$ is the Euler-Mascheroni
constant.

By means of the asymptotic expansion of $K_{\pm}(t)$ for $t\mapsto 1$, which
is equivalent to expansion of $H_{\pm}(z)$ for small z and using the formulae
(B.4) and (B.5) one arrives at complete asymptotic of $r_{\pm}(N)$:

\begin{theorem}

For $N\mapsto \infty$ one has
$$
r_{\pm}(N)=C_{\pm}(p)N^{(2q Z_p(0)-p-2)/(2(1+p))}\times
$$
$$
\exp\left\{\frac{1+p}{p}[q A\Gamma(1+p)\zeta_{\pm}(1+p)]^
{1/(1+p)}N^{p/(1+p)}\right\}[1+O(N^{-\kappa_{\pm}})]\mbox{,}
\eqno{(B.12)}
$$
$$
C_{\pm}(p)=[q A\Gamma(1+p)\zeta_{\pm}(1+p)]^{(1-2q Z_p(0))/(2p+2)}
\frac{\exp(q Z_p'(0))}{[2\pi(1+p)]^{1/2}}\mbox{,}
\eqno{(B.13)}
$$
$$
\kappa_{\pm}=\frac{p}{1+p}\min \left(\frac{C_{\pm}}{p}-\frac{\delta}{4},
\frac{1}{2}-\delta\right)\mbox{,}
$$
and $0<\delta<\frac{2}{3}$.
\end{theorem}

\s{Appendix C. Free Energy of Fundamental (Super) p-Brane}

The one-loop free energy of fields containing in (super) $p$-brane
 can be evaluated substituting the mass operators
(A.1) and (A.2) and making use the Mellin-Barnes representation
\cite{byts93-394-423,byts96-266-1}. Free energies have the form
$$
{\cal F}_{p}(\beta)=-\frac{(4\pi)^{-\frac{D}{2}}}{4\pi i}\int_{Re s=c}ds
\zeta_{-}(s)\left(\frac{\beta}{2}\right)^{-s}\Gamma\left(\frac{s}{2}\right)
\Gamma\left(\frac{s-D}{2}\right)\Tr[M_{p}^2]^{\frac{D-s}{2}}\mbox{,}
\eqno{(C.1)}
$$
$$
{\cal F}_{sp}(\beta)=-\frac{(4\pi)^{-\frac{D}{2}}}{2\pi i}\int_{Re s=c'}ds
\zeta_{+}(s)\left(\frac{\beta}{2}\right)^{-s}\Gamma\left(\frac{s}{2}
\right)\Gamma\left(\frac{s-D}{2}\right)
\Tr[M_{sp}^2]^{\frac{D-s}{2}}\mbox{,}
\eqno{(C.2)}
$$
where
$$
\Tr[M_{p}^2]^{\frac{D-s}{2}}=\left[\Gamma\left(\frac{s-D}{2}\right)\right]^{-1}
\int_0^\infty dtt^{\frac{s-D}{2}-1}H_{-}(t)\mbox{,}
\eqno{(C.3)}
$$
$$
\Tr[M_{sp}^2]^{\frac{D-s}{2}}=\left[\Gamma\left(\frac{s-D}{2}\right)\right]^{-1}
\int_0^\infty dtt^{\frac{s-D}{2}-1}H_{+}(t)H_{-}(t)\mbox{,}
\eqno{(C.4)}
$$
and
$$
H_{\pm}(t)=\prod_{\vec n\in \Z^p/\{0\}}
\left[1\pm
\exp\left(-t\omega_{\vec n}\right)\right]^{\pm(D-p-1)}\mbox{.}
\eqno{(C.5)}
$$
Using  Eqs.(B.4) and (B.5) of Proposition 1 one can obtain for
$t\mapsto 0$,
$$
H_{-}(t)=t^{\frac{q}{2}}\exp\left[V_{-}^{2p}(p)t^{-p}+U(p)+O\left(t^{c_{-}}
\right)\right]\mbox{,}
\eqno{(C.6)}
$$
$$
H_{+}(t)H_{-}(t)=\left(\frac{t}{2}\right)^{\frac{q}{2}}\exp\left[V_{+}^{2p}(p)t^{-p}+
U(p)+O\left(t^{c_{+}}\right)\right]\mbox{,}
\eqno{(C.7)}
$$
where $q\equiv D-p-1$ and
$$
V_{-}(p)=\left[q 2^{p-2}\pi^{\frac{p-1}{2}}\Gamma\left(\frac{1+p}{2}\right)
\zeta_{-}(1+p)\right]^{\frac{1}{2p}}\mbox{,}
\eqno{(C.8)}
$$
$$
V_{+}(p)=\left[q 2^{p-2}\pi^{\frac{p-1}{2}}\Gamma\left(\frac{1+p}{2}\right)
\left(\zeta_{-}(1+p)+\zeta_{+}(1+p)\right)\right]^{\frac{1}{2p}}\mbox{,}
\eqno{(C.9)}
$$
$$
U(p)=\frac{q}{2}\left[\frac{\Gamma(\frac{p}{2})}{\pi p}B_{p}+\frac{1}{2}
\psi(\frac{p}{2})-\mbox{log}\pi-\frac{1}{2}\gamma\right]\mbox{.}
\eqno{(C.10)}
$$

It is convenient to use the same substruction procedure as in Ref.
\cite{acto93-315-74} for the divergent term in $H_{\pm}$ in order to
regularize the free energy integrals,
$$
\Tr[M^2]^{\frac{D-s}{2}}=C_{\pm}(p)\frac{\Gamma\left(\frac{p-s+1}{2p}
\right)}{\Gamma\left(\frac{s-D}{2}\right)}V_{\pm}^s(p)\Re(-1)^{(s-p-1)/(2p)}+
\frac{1}{\Gamma\left(\frac{s-D}{2}\right)}G_{\pm}(s;\mu)\mbox{,}
\eqno{(C.11)}
$$
where
$$
C_{-}(p)=\frac{1}{p}\exp[U(p)]V_{-}^{-p-1}(p)\mbox{,}
\eqno{(C.12)}
$$
$$
C_{+}(p)=\frac{1}{p}2^{-\frac{q}{2}}\exp[U(p)]V_{+}^{-p-1}(p)\mbox{,}
\eqno{(C.13)}
$$
$$
G_{-}(s;\mu)=\int_{0}^{\mu}dtt^{(s-D)/2-1}\left\{H_{-}(t)-t^{q/2}
\exp[V_{-}^{2p}(p)t^{-p}+U(p)]\right\}\mbox{,}
\eqno{(C.14)}
$$
$$
G_{+}(s;\mu)=\int_{0}^{\mu}dtt^{(s-D)/2}\left\{H_{+}(t)H_{-}(t)-
\left(\frac{t}{2}\right)^{q/2}\exp[V_{+}^{2p}(p)t^{-p}+U(p)]\right\}\mbox{.}
\eqno{(C.15)}
$$
In Eqs. (C.14) and (C.15) an infrared cutoff parameter $\mu$ has been
introduced. It should be noted that in contrast with the supersymmetric case
the regularization in the infrared region of the integral defining the
analytic function $G_{-}(s;\mu)$ cannot be removed (there are tachyons in the
$p$-brane spectrum).

The one-loop free energy can be written as follows
$$
{\cal F}_{p}(\beta)=-\frac{(4\pi)^{-\frac{D}{2}}}{4\pi i}\int_{\Re s=c}ds
[\Phi_{-}(s)+\Xi_{-}(s)]\mbox{,}
\eqno{(C.16)}
$$
$$
{\cal F}_{sp}(\beta)=-\frac{(4\pi)^{-\frac{D}{2}}}{2\pi i}\int_{\Re s=c'}ds
[\Phi_{+}(s)+\Xi_{+}(s)]\mbox{,}
\eqno{(C.17)}
$$
where
$$
\Phi_{\pm}(s)=C_{\pm}x_{\pm}^s\Re(-1)^{(s-p-1)/(2p)}\Gamma\left(\frac{s}{2}
\right)\Gamma\left(\frac{p+1-s}{2p}\right)\zeta_{\pm}(s)\mbox{,}
\eqno{(C.18)}
$$
$$
\Xi_{\pm}(s)=G_{\pm}(s;\mu)\left(\frac{\beta}{2}\right)^{-s}
\Gamma(\frac{s}{2})\zeta_{\pm}(s)\mbox{,}
\eqno{(C.19)}
$$
and $x_{\pm}=2V_{\pm}(p)\beta^{-1}$. The various residues of the meromorphic
functions $\Phi_{\pm}(s)$ and $\Xi_{\pm}(s)$ are
$$
\Res\left[\Phi_{\pm}(s), s=p(2k-1)+1)\right]=C_{\pm}(p)\frac{\Gamma
\left(pk+\frac{1-p}{2}\right)}{\Gamma(k)}\zeta_{\pm}(2pk+1-p)
x^{1+p(2k-1)}\mbox{,}
$$
$$
\Res\left[\Phi_{-}(s), s=0\right]=-C_{-}(p)\Gamma\left(\frac{1+p}{2p}
\right)\Re(-1)^{-(1+p)/(2p)}\mbox{,}
$$
$$
\Res\left[\Xi_{\pm}(s),s=1\right]=\sqrt{\pi}2^{\mp 1}G_{\pm}(1;\mu)
\beta^{-1}\mbox{,}
$$
$$
\Res\left[\Xi_{-}(s),s=0\right]=-G_{-}(0;\mu)\mbox{.}
$$

Finally a formal computation of the integrals (C.16) and (C.17) gives
$$
{\cal F}_{p}(\beta)=-\frac{1}{2(4\pi)^{D/2}}\left\{C_{-}(p)\sum_{k=1}
^{\infty}\frac{\Gamma\left(pk+\frac{1-p}{2}\right)}{\Gamma(k)}
\zeta_{R}(2pk+1-p)x_{-}^{1+p(2k-1)}+\right.
$$
$$
\left.2\sqrt{\pi}G_{-}(1;\mu)\beta^{-1}-G_{-}(0;\mu)-C_{-}(p)\Gamma
\left(\frac{1+p}{2p}\right)\Re(-1)^{-(1+p)/(2p)}\right\}+
$$
$$
{\cal F}_{p{\cal R}}(x)+{\cal F}_{p{\cal R}}(x;\mu)\mbox{,}
\eqno{(C.20)}
$$
$$
{\cal F}_{sp}(\beta)=-\frac{1}{(4\pi)^{D/2}}\left\{C_{+}(p)\sum_{k=1}
^{\infty}\frac{\Gamma\left(pk+\frac{1-p}{2}\right)}{\Gamma(k)}
\zeta_{+}(2pk+1-p)x_{+}^{1+p(2k-1)}+\right.
$$
$$
\left.\frac{1}{2}\sqrt{\pi}G_{+}(1;\mu)\beta^{-1}\right\}+{\cal F}_
{sp{\cal R}}(x)+{\cal F}_{sp{\cal R}}(x;\mu)\mbox{,}
\eqno{(C.21)}
$$
where ${\cal F}_{p{\cal R}}(x)$,\, ${\cal F}_{sp{\cal R}}(x)$ and ${\cal F}_
{p{\cal R}}(x;\mu)$,\, ${\cal F}_{sp{\cal R}}(x;\mu)$ are the contributions
$\Phi_{\pm}(s)$ and $\Xi_{\pm}(s)$ along the arc of radius ${\cal R}$ in the
right half-plane. If $|x|<1$ then ${\cal F}_{p{\cal R}}(x)$ and ${\cal F}_
{sp{\cal R}}(x)$ vanishes when ${\cal R}\mapsto\infty$.

\bf Strings\rm

Returning to the string case we note that for $p=1$ the serie in Eqs. (C.20)
and (C.21) can be evaluated in the closed forms \cite{byts96-266-1} ,\cite{byeo93}.The sums
of these serie for open string and superstring (without gauge group) have
respectively the forms
$$
\sum_{k=1}^{\infty}\zeta_{R}(2k)x_{-}^{2k}=\frac{1}{2}-\frac{1}{2}
\cot(\pi x)\mbox{,}
\eqno{(C.22)}
$$
$$
\sum_{k=1}^{\infty}\zeta_{R}(2k)\left(1-2^{-2k}\right)x_{+}^{2k}=
\frac{\pi x}{4}\tan\left(\frac{\pi x}{2}\right)\mbox{.}
\eqno{(C.23)}
$$
The finite radius of Laurent series convergence $|x_{\pm}|<1$ corresponds to
the critical temperature in string thermodynamics: $x_{\pm}=\beta_{\pm}^c/
\beta$ and $\beta_{\pm}^c=2V_{\pm}(1)$. As a result $\beta_{-}^{c}=\sqrt{8}
\pi$ and $\beta_{+}^{c}=2\pi$ (review of string theory at non-zero temperature
may be found in Ref. \cite{od92-15-1}). Written in trigonometric form
Expressions (C.22) and (C.23) displays a certain periodicity in temperature,
the physical meaning of which is still obscure. For the both open bosonic and
supersymmetric strings the $\beta$-behavior of the free energy (C.22), (C.23)
has the dependence on temperature near the Hagedorn transition which looks like
one found in Ref. \cite{atic88-310-291}.

\bf p-Branes\rm

For $p>0$ the power series (C.20) and (C.21) are divergent for any $x_{\pm}>0$.
Nevertheless one can construct an analytic continuation of these expressions.
Let us define for $|z|<\infty$ two serie
$$
{\cal J}_{\pm}(z)=\sum_{k=0}^{\infty}\frac{\sqrt{\pi}}{\Gamma(k+1)
\Gamma\left(pk+\frac{p+2}{2}\right)}\nu_{\pm}(k;p)\left(\frac{z}{2}\right)^
{p(2k+1)+1}\mbox{.}
\eqno{(C.24)}
$$
In addition the factors $\nu_{\pm}(k;p)$ have the form
$$
\nu_{-}(k;p)=(-1)^{pk+1}\mbox{,}
\eqno{(C.25)}
$$
$$
\nu_{+}(k;p)=\nu_{-}(k;p)\left[1-2^{-p(2k+1)-1}\right]\mbox{.}
\eqno{(C.26)}
$$
For finite variable $z$ these serie converge and convergence is improving rapidly
with increasing integer number $p$. Let $z_{\pm}=j\cdot2\pi x_{\pm}$, then for
the serie
$$
\sum_{j=1}^{\infty}{\cal J}_{\pm}(j\cdot2\pi x_{\pm})=\sum_{j=1}^{\infty}
\sum_{k=0}^{\infty}\frac{\sqrt{\pi}}{\Gamma(k+1)\Gamma\left(pk+\frac{p+2}{2}
\right)}\nu_{\pm}(k;p)(j\pi x_{\pm})^{p(2k+1)+1}\mbox{,}
\eqno{(C.27)}
$$
one can commute the (now divergent) sum $\Sigma_j$ through $\Sigma_k$ which
generates extra terms on the right $J_{\pm}(x,p)$. Thus the result is
$$
\sum_{j=1}^{\infty}{\cal J}_{\pm}(j2\pi x_{\pm}) + J_{\pm}(x,p)
$$
$$
=\sum_{k=0}^{\infty}\frac{\sqrt{\pi}}{\Gamma(k+1)\Gamma\left(
pk+\frac{p+2}{2}\right)}\nu_{\pm}(k;p)\zeta_{R}[-p(2k+1)](\pi x_{\pm})^
{p(2k+1)+1}
$$
$$
=\sin\left(\frac{\pi p}{2}\right)\sum_{k=1}^{\infty}
\frac{\Gamma\left(pk+\frac{1-p}{2}\right)}{\Gamma(k)}\zeta_{\pm}(2pk+1-p)
x_{\pm}^{p(2k-1)+1}\mbox{,}
\eqno{(C.28)}
$$
where, for example, $J_{-}(x,p)=\pi\left[px\Gamma\left(\frac{(p-1)(p+1)}{2p}
\right)\right]^{-1}$ (see Ref. \cite{acto93-315-74}), and in the second
equality the functional equation for $\zeta_{R}(s)$ is used. The vanishing of
the factor $\sin(p\pi/2)$ on the right hand side for even $p$ merely expresses
the vanishing of $\zeta_{R}[-p(2k+1)]$ in the first equality. For $p=2k, k\in
\Z_{+}$, the right hand side of Eq. (C.28) is zero and the function (C.27)
therefore sums to a single term.


\begin{thebibliography}{10}

\bibitem{stro96-379-99}
A. Strominger and C. Vafa, Phys. Lett. {\bf B 379}, 99 (1996).

\bibitem{das96u-52}
S.R. Das and S. Mathur, {\em "Excitations of D-Strings, Entropy and Duality"},
hep-th/9601152 (1996).

\bibitem{call96-475-645}
C. Callan and J. Maldacena, Nucl. Phys. {\bf B 475}, 645 (1996).

\bibitem{horo96-77-2368}
G.T. Horowitz and A. Strominger, Phys. Rev. Lett. {\bf 77}, 2368 (1996).

\bibitem{ghos96u-57}
A. Ghosh and P. Mitra, {\em "Entropy of Extremal Dyonic Black Holes"},
hep-th/9602057 (1996).

\bibitem{beck96u-65}
J. Beckenridge, R. Meyrs, A. Peet and C. Vafa, {\em "D-Branes and Spinning
Black Holes"}, hep-th/9602065 (1996).

\bibitem{gubs96u-35}
S.S. Gubser, I.R. Klebanov and A.W. Peet, {\em "Entropy and Temperature of
Black 3-Branes"}, hep-th/9602135 (1996).

\bibitem{das96u-72}
S.R. Das, {\em "Black Hole Entropy and String Theory"}, hep-th/9602172 (1996).

\bibitem{beck96-381-423}
J. Beckenridge, D.A. Lowe, R. Myers, A. Peet, A. Strominger and C. Vafa,
Phys. Lett. {\bf B 381}, 423 (1996).

\bibitem{mald96-77-428}
J. Maldacena and A. Strominger, Phys. Rev. Lett. {\bf 77}, 428 (1996).

\bibitem{john96-378-78}
C.V. Johnson, R.R. Khuri and R.C. Myers, Phys. Lett. {\bf B 378}, 78 (1996).

\bibitem{horo96-383-151}
G.T. Horowitz, J. Maldacena and A. Strominger, Phys. Lett. {\bf B 383}, 151
(1996).

\bibitem{dijk96u-26}
R. Dijkgraaf, E. Verlinde and H. Verlinde, {\em "BPS Spectrum of the Five-
Brane and Black Hole Entropy"}, hep-th/9603126 (1996).

\bibitem{kleb96u-89}
I.R. Klebanov and A.A.Tseytlin, {\em "Entropy of Near-Extremal Black p-Branes"},
hep-th/9604089 (1996).

\bibitem{lu96u-27}
H. L{\"u}, S. Mukherji, C.N. Pope and J. Rahmfeld, {\em "Loop-Corrected
Entropy of Near-Extremal Dilatonic p-Branes"}, hep-th/9604127 (1996).

\bibitem{duff91-253-113}
M.J. Duff and K.S. Stelle, Phys. Lett. {\bf B 253}, 113 (1991).

\bibitem{guve92-276-49}
R. G{\"u}ven, Phys. Lett. {\bf B 276}, 49 (1992).

\bibitem{horo91-360-197}
G.T. Horowitz and A. Strominger, Nucl. Phys. {\bf B 360}, 197 (1991).

\bibitem{duff91-273-409}
M.J. Duff and J.X. Lu, Phys. Lett. {\bf B 273}, 409 (1991).

\bibitem{duff95-273-479}
M.J. Duff, S. Ferrara, R.R. Khuri and J. Rahmfeld, Phys. Lett.
{\bf B273}, 479 (1995).

\bibitem{lu95u-53}
H. L{\"u} and C.N. Pope, {\em "Multi-Scalar p-Brane Solitons"}, hep-th/9512153
(1995).

\bibitem{haly96u-68}
E. Halyo, {\em "Reissner-Nordstr\o{m} Black Holes and Strings with Rescaled
Tension"}, hep-th/9610068 (1996).

\bibitem{lu95-456-669}
H. L{\"u}, C.N. Pope, E. Sezgin and K.S. Stelle, Nucl. Phys. {\bf B 456},
669 (1995).

\bibitem{cvet96-53-584}
M. Cveti\v{c} and D. Youm, Phys. Rev. {\bf B 53}, 584 (1996).

\bibitem{duff95-345-441}
M.J. Duff and J. Rahmfeld, Phys. Lett. {\bf B 345}, 441 (1995).

\bibitem{sen95-440-421}
A. Sen, Nucl. Phys. {\bf B 440}, 421 (1995).

\bibitem{sen95-10-2081}
A. Sen, Mod. Phys. Lett. {\bf A 10}, 2081 (1995).

\bibitem{duff96-459-125}
M.J. Duff, J.T. Liu and J. Rahmfeld, Nucl. Phys. {\bf B 459}, 125 (1996).

\bibitem{rahm95u-89}
J. Rahmfeld, {\em "Extremal Black Holes as Bound States"}, hep-th/9512089
(1995).

\bibitem{lu96-465-127}
H. L{\"u} and C.N. Pope, Nucl. Phys. {\bf B 465}, 127 (1996).

\bibitem{lu96u-26}
H. L{\"u} and C.N. Pope, {\em "Black p-Branes and Their Vertical Dimensional
Reduction"}, hep-th/9609126 (1996).

\bibitem{duff96u-52}
M.J. Duff, H. L{\"u} and C.N. Pope, {\em "The Black p-Branes of M-Theory"},
hep-th/9604052 (1996).

\bibitem{lu96u-07}
H. L{\"u}, S. Mukherji and C.N. Pope, {\em "Cosmological Solutions in String
Theories"}, hep-th/9610107 (1996).

\bibitem{luka96u-38}
A. Lukas, B.A. Ovrut and D. Waldram, {\em "String and M-Theory Cosmological
Solutions with Ramond Forms"}, hep-th/9610238 (1996).

\bibitem{dabh90-340-33}
A. Dabholkar, G.W. Gibbons, J.A. Harvey and F. Ruiz Ruiz, Nucl. Phys.
{\bf B 340}, 33 (1990).

\bibitem{call91-359-611}
G.G. Callan, J.A. Harvey and A. Strominger, Nucl. Phys. {\bf B 359},
611 (1991).

\bibitem{call91-367-60}
G.G. Callan, J.A. Harvey and A. Strominger, Nucl. Phys. {\bf B 367},
60 (1991).

\bibitem{duff94-416-301}
M.J. Duff and J.X. Lu, Nucl. Phys. {\bf B 416}, 301 (1994).

\bibitem{byts93-304-235}
A.A. Bytsenko, K. Kirsten and S. Zerbini, Phys. Lett. {\bf B 304}, 235 (1993).

\bibitem{byts94-9-1569}
A.A. Bytsenko, K. Kirsten and S. Zerbini, Mod. Phys. Lett. {\bf A 9},
1569 (1994).

\bibitem{byts96-266-1}
A.A. Bytsenko, G. Cognola, L. Vanzo and S. Zerbini, Phys. Reports {\bf 266},
1 (1996).

\bibitem{byod90}
A.A. Bytsenko and S.D. Odintsov, Phys. Lett. {\bf B 243}, 63 (1990);
{\bf B 245}, 21 (1990).

\bibitem{alva91-43-3990}
E. Alvarez, T. Ortin and M.A.R. Osorio, Phys. Rev. {\bf D 43}, 3990 (1991).

\bibitem{alva92-7-2889}
E. Alvarez and T. Ortin, Mod. Phys. Lett. {\bf A 7}, 2889 (1992).

\bibitem{acto93-315-74}
A.A. Actor and A.A. Bytsenko, Phys. Lett. {\bf B 315}, 74 (1993).

\bibitem{gibb95-12-297}
G.W. Gibbons, G.T. Horowitz and P.K. Townsend, Class. Quant. Grav. {\bf 12},
297 (1995).

\bibitem{borc95-120-161}
R.E. Borcherds, Invent. Math. {\bf 120}, 161 (1995).

\bibitem{harv96-463-315}
J.A. Harvey and G. Moore, Nucl. Phys. {\bf B 463}, 315 (1996).

\bibitem{mald96-475-679}
J. Maldacena and L. Susskind, Nucl. Phys. {\bf B 475}, 679 (1996).

\bibitem{haly96u-12}
E. Halyo, A. Rajaraman and L. Susskind, {\em "Braneless Black Holes"},
hep-th/9605112 (1996).

\bibitem{schwarz95}
J.H.Schwarz, Phys. Lett. {\bf B 360}, 13 (1995) (E: {\bf B 364}, 252 (1995)).

\bibitem{russo96}
J.G. Russo and A.A. Tseytlin, {\em "Waves, Boosted Branes and BPS States
in M-Theory"}, hep-th 9611047  (1996).

\bibitem{berg87-185-330}
E. Bergshoeff, E. Sezgin and P.K. Townsend, Ann. Phys. {\bf 185}, 330 (1987).

\bibitem{duff88-297-515}
M.J. Duff, T. Inami, C.N. Pope, E. Sezgin and K. Stelle, Nucl. Phys.
{\bf B 297}, 515 (1988).

\bibitem{mein54-59-338}
G. Meinardus, Math. Z. {\bf 59}, 338 (1954).

\bibitem{mein54-61-289}
G. Meinardus, Math. Z. {\bf 61}, 289 (1954).

\bibitem{andr76b}
G.E. Andrews, {\em "The Theory of Partitions"}. In Encyclopedia of Mathematics
and its Applications, Addison-Wesley Publishing Company (1976).

\bibitem{byts93-394-423}
A.A. Bytsenko, E. Elizalde, S.D. Odintsov and S. Zerbini, Nucl. Phys.
{\bf B 394}, 423 (1993).

\bibitem{od92-15-1}
S.D. Odintsov, Rivista Nuovo Cim. {\bf 15}, 1 (1992).

\bibitem{atic88-310-291}
J.J. Atick and E. Witten, Nucl. Phys. {\bf B 310}, 291 (1988).

\bibitem{tm97}
T. Muto, Phys. Lett. {\bf B 391},310 (1997).

\end{thebibliography}
\end{document}